\documentclass[10pt,aps,twocolumn, prb, superscriptaddress, floatfix]{revtex4-1}
\usepackage[T1]{fontenc}
\usepackage[latin1]{inputenc}
\usepackage[english]{babel}
\usepackage{graphicx}
\usepackage{paralist}
\usepackage{array}
\usepackage{amsmath}
\usepackage{amssymb}
\usepackage{ragged2e}
\usepackage{dsfont}
\usepackage{booktabs}
\usepackage{url}
\usepackage{color}
\usepackage{hyperref}

\newcommand{\bbra}[1]{\langle\langle #1|}
\newcommand{\kket}[1]{|#1\rangle\rangle}
\newcommand{\bbrakket}[2]{\langle\langle #1|#2 \rangle\rangle}
\newcommand{\aav}[1]{\langle\langle #1 \rangle\rangle}

\newcommand{\beq}{\begin{eqnarray}}
\newcommand{\eeq}{\end{eqnarray}}
\newcommand{\rb}[1]{\left( #1 \right)}
\newcommand{\ew}[1]{\langle #1 \rangle}
\newcommand{\eq}[1]{Eq.~(\ref{#1})}

\newcommand{\eww}[1]{\langle\! \langle #1\rangle\! \rangle}

\begin{document}
\title{Full counting statistics of random transition-rate matrices}
\author{Uliana Mordovina}
\affiliation{Institut f\"ur Theoretische Physik, Hardenbergstrasse 36, TU Berlin, D-10623 Berlin, Germany}
\author{Clive Emary}
\affiliation{Department of Physics and Mathematics, University of Hull, Kingston-upon-Hull, HU6 7RX, United Kingdom}
\begin{abstract}
We study the full counting statistics of current of large open systems through the application of random matrix theory to transition-rate matrices.  We develop a method for calculating the ensemble-averaged current-cumulant generating functions based on an expansion in terms of the inverse system size.  We investigate how different symmetry properties and different counting schemes affect the results.
\end{abstract}

\maketitle

\section{Introduction}

Since its application to nuclear spectra in the 1950s \cite{wigner1957proceedings,*wigner1955characteristic,dyson:140,*dyson:157,*dyson:166}, random matrix theory has provided a way of treating large, complicated or chaotic physical systems for which a detailed microscopic description is impossible, undesirable or unilluminating \cite{Beenakker1997,Guhr1998,Forrester_2003,Meh2004}.
The matrices considered in random matrix theory are typically Hermitian (or unitary in the case of e.g. scattering matrices 
\cite{Beenakker1997}), but non-Hermitian matrices have also attracted some attention, 
e.g.~Refs.~\cite{Ginibre1965,Bray1988,Fyodorov1999,Staring2003,Amir2008}.
In particular, in Ref.~\cite{PhysRevE.80.021140} Timm considered random transition-rate matrices 
(we shall use the terminology {\em Liouvillian} here), which are non-Hermitian matrices arising from 
the description of a dynamical process in terms of a rate or master equation.  Timm gave a detailed account of the spectral 
properties of such matrices.

Rate equations find application across the sciences \cite{vanKampenbook,Gardinerbook}, but we are particularly interested
here in their application in quantum transport, where the Pauli master equation is used to describe charge flow through a small 
system or device, such as a quantum dot or molecule, that is  weakly coupled to the leads.
Beyond just looking at average populations and current  flows,
the master equation approach lends itself well to the calculation of the so-called full counting statistics 
(FCS) \cite{LEV93, LEV96, Bagrets2003}, i.e. the probability distribution function of the numbers of charges 
transferred through the device in a given time interval.  This information is usually encapsulated as a cumulant generating 
function (CGF) from which, not only  the mean currents and their fluctuations (shotnoise) can be calculated, but also the higher 
current cumulants, which give further insight into the transport process.
Following the work of Bagrets and Nazarov \cite{Bagrets2003}, the theory of FCS within the master equation approach has been extensively developed, see for example Refs.~\cite{Emary2007d,*Marcos2010a,Flindt2008, Emary2009b,Flindt2010}, and many results from this approach for quantum-dot systems have been confirmed experimentally \cite{Fujisawa16062006, PhysRevLett.96.076605, PhysRevB.75.075314, Flindt23062009, Sukhorukov2007}.

In this paper we use these techniques to calculate the FCS of ensembles of master equation with random Liouvillian matrices as introduced in Ref.~\cite{PhysRevE.80.021140}.  Using a perturbative approach, we calculate  the ensemble-averaged CGF as a series in the inverse of system size $N$.
Given a particular Liouvillian, the FCS is not unique.  Rather, the FCS depends on the details of what is being counted and how.  In this work we will mainly consider the simplest Ansatz in this respect and count the number of times the system undergoes a single particular transition from the many available to it.
We will consider both bi- and uni-directional counting at this transition and also discuss how the results depend on the symmetry of the underlying rate equation.

As we set out below (see Fig.~\ref{fig:network}), the dynamics of the rate equation here can be visualised as the diffusion 
of a particle on a network, see e.g. \cite{Bray1988,Jespersen2000,Almaas2003,Noh2004,Lopez2005}, in which every node is attached to all the others and transitions between the nodes occur at rate given by the off-diagonal constituents of the Liouvillian. In this light, then, we are calculating here the FCS of a network where we count every time 
the random walker moves between two particular nodes of the network.

The paper is structured as follows: In Sec.~\ref{sec:model} we describe the master equation and random Liouvillians, as well as define our approach to counting in these systems.  In Sec.~\ref{sec:expansion} we provide the building blocks for our expansion in the limit of large system size, and in Sec.~\ref{sec:CGFs} we give our expressions for the ensemble-averaged FCS.  We then present some concluding remarks in Sec.~\ref{SEC:disc}.

\section{Model}
\label{sec:model}

We begin with the master equation for the system:
\begin{equation}
 \cfrac{\mathrm{d}}{\mathrm{d} t} \kket{\rho(t)}=\mathcal{L}\kket{\rho(t)},
 \label{ME}
\end{equation}
where $\kket{\rho(t)}$ denotes the probability vector that contains the probabilities of finding the system in each of its accessible states at a given time, and where $\mathcal{L}$ is the transition rate matrix or {\em Liouvillian} superoperator in matrix representation for the system
\footnote{This notation is useful as it admits immediate generalisation to ``quantum-optics style'' quantum master equations, where $\mathcal{L}$ is of a more general Lindblad form that also contains quantum-coherent contributions}.  Off-diagonal elements of $\mathcal{L}$ are the transition rates $\Gamma_{ij}$ between the various system states and the diagonal elements follow from the conservation of probability:
\begin{equation}
\mathcal{L}_{ij}= \begin{cases}
           \Gamma_{ij} &  i \neq j,\\
            -\sum\limits_{ m \neq i}{\Gamma_{im}} &  i = j.
                                        \end{cases}
\label{eq:L}
\end{equation}
The stationary state of the system (assumed unique), which we denote as $\kket{\rho_0}$, may be found as the right eigenvector of Liouvillian:  $\mathcal{L}\kket{\rho_0} = 0$.  The left nullvector, $\bbra{\widetilde{\rho}_0} \mathcal{L} = 0$ is
\beq
  \bbra{\widetilde{\rho}_0} 
  = \bbra{\mathds{1}} 
  \equiv (1,\ldots,1)
  \label{eq:TrDVec}
  ,
\eeq
irrespective of the values of the rates $\Gamma_{ij}$. We normalise the stationary state such that $\eww{\widetilde{\rho}_0|\rho_0} = 1$.

\begin{figure}
  \includegraphics[width=.7\linewidth]{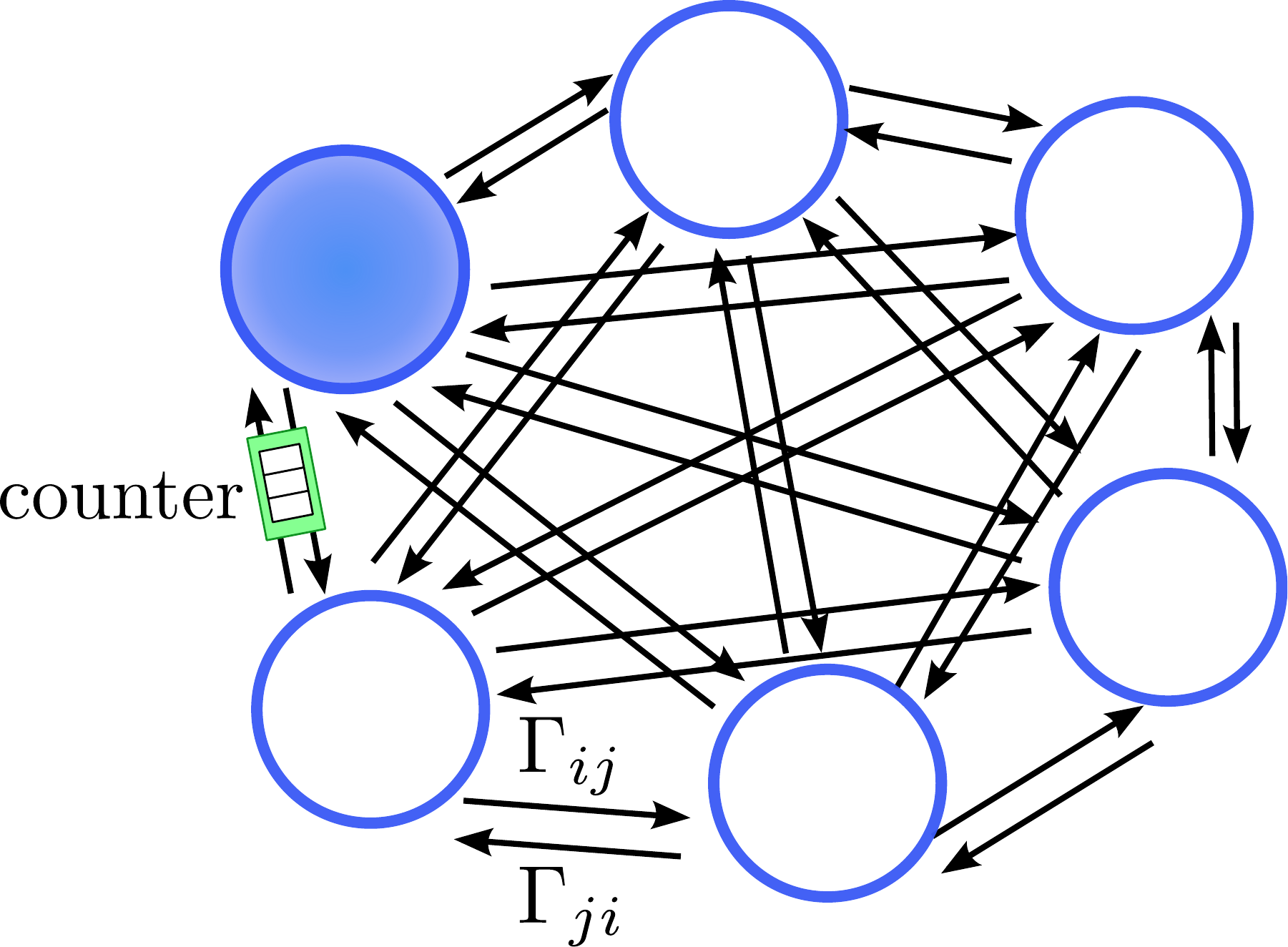}
  \caption{ 
    The master equation studied here can be thought as describing the diffusion of a particle on a network where each node is 
    coupled to every other one and a transition from node $i$ to node $j$ occurs with rate $\Gamma_{ij}$.
    Here we sketch such a network with $N=6$ nodes. 
    The occupied node is shown as a filled circle, the unoccupied nodes as empty. 
    To obtain the full counting statistics (FCS), a counter is inserted between two of the nodes,  which counts every time the system moves between the two nodes.
    }
 \label{fig:network}
\end{figure}

The system can be visualized by a network (Fig. \ref{fig:network}) in which each node represents a single system state. 
With a fully populated Liouvillian, each node is connected to every other node in the network. The dynamics described by \eq{ME} 
can be thought as a diffusion on this network, with transitions between nodes taking place with rates $\Gamma_{ij}$.

Following Ref.~\cite{PhysRevE.80.021140} we treat matrix $\mathcal{L}$ as an element of a random-matrix ensemble, where the  off-diagonal elements are chosen from an appropriate non-negative, real distribution. 
We will consider continuous random variables with density $f_{\Gamma_{ij}}(x)$, such that $\textstyle \int_a^b f_{\Gamma_{ij}}(x) \mathrm{d}x$ is the probability that the rate $\Gamma_{ij}$ takes a value in the interval $[a,b]$  
In this work we will consider as examples the exponential distribution with $f_{\Gamma_{ij}}(x)=\frac{1}{\overline{\Gamma}}e^{-x/\overline{\Gamma}}$
and moments $\overline{\Gamma^n}=n!\overline{\Gamma}^n$, 
as well as the  more general gamma-distribution
$f_{\Gamma_{ij}}(x)=\frac{e^{-x/\beta } \beta ^{-\alpha } x^{\alpha -1}}{\Gamma(\alpha)}$
with moments $\overline{\Gamma^n}=\frac{(\alpha+n-1)!}{\Gamma(\alpha)}\beta^n$, where $\Gamma(\alpha)$ is the gamma function.

The ensemble of random-rate matrices is defined by the structure (Eq.~(\ref{eq:L})), by the choice of $\Gamma_{ij}$-distribution, and also by any further symmetry conditions.  In this latter respect, we will consider both symmetric matrices, i.e. with $\Gamma_{ij}=\Gamma_{ji}$,
as well as asymmetric matrices where the rates $\Gamma_{ij}$ and $\Gamma_{ji}$ chosen independently of one another.

\subsection{FCS}
In the standard transport context, FCS concerns itself with the probability  $P(n;t)$  that $n$ electrons are transferred through a conductor
to a given lead in time $t$ \cite{LEV93, LEV96}.  In the mater equation picture, this translates into counting the number of ``jumps''
the system makes as electrons enter and leave the lead \cite{Bagrets2003}.  In this setting, the FCS are calculated by first dividing the
Liouvillian $\mathcal{L}$ into parts that correspond to jump processes to be counted and the rest.  Given a particular Liouvillian,
this process is not unique, since there are may different ways in which we can count the various processes.

In considering the FCS of the random Liouvillians, we will consider the simplest counting, as is best illustrated by studying the network picture of Fig.~\ref{fig:network}.
The occupied node diffuses around the network. On one of the vertices of the network, we place a counter such that every time 
the network switches between the two states connected by the counter, we increment the counter variable $n$ for one way 
and decrement it if the system moves in the other. As the dynamics are stochastic, this counting procedure leads to a probability $P(n;t)$ for $n$ such transitions to have occurred in time $t$. This quantity, or its Fourier transform, the CGF, is the quantity in which we are interested.  This counting is ``bidirectional'' and we will develop the FCS notation below in terms of it. It is also possible (simpler even) to describe  uni-directional counting, and we will return to this at the end of the section.

With bidrectional counting, we split Liouvillian as
\beq
 \mathcal{L} = \mathcal{L}_0 + \mathcal{J}^++ \mathcal{J}^-
 ,
\eeq
where $\mathcal{J}^+$ and $\mathcal{J}^-$ are jump (super-)operators describing a single transition between states in the forward and backward directions respectively.
For our purposes here, it does not matter between which states  we situate the counter and thus we choose states $i=1$ and $j=N$ as the relevant states. 
This leads to following matrix representations for the jump operators
\begin{align}
 \mathcal{J}^+_{ij} & =\Gamma_{ij} \delta_{1i}\delta_{Nj};
 \label{eq:j+} \\
 \mathcal{J}^-_{ij} & =\Gamma_{ij}\delta_{Ni}\delta_{1j}.
 \label{eq:j-}
\end{align}

With this partitioning of the Liouvillian, we can obtain information about counting variable $n$ by writing down the corresponding  system 
of $n$-resolved master equations \cite{Cook1981,Gurvitz1996,*Gurvitz1998}
\beq
  \cfrac{\mathrm{d}}{\mathrm{d} t} \kket{\rho^{(n)}(t)} &=& 
  \mathcal{L}_0\kket{\rho^{(n)}(t)}
  +\mathcal{J}^+\kket{\rho^{(n-1)}(t)}
  \nonumber\\
  &&  
  +\mathcal{J}^-\kket{\rho^{(n+1)}(t)},
  \label{eq:n_alg}
\eeq
where $\kket{\rho^{(n)}}$ is the state of the system conditioned on $n$ counting events having taken place.  The probability $P(n;t)$ is then given by 
\begin{equation}
 P(n;t)=\mathrm{Tr}\left\{\kket{\rho^{(n)}(t)}\right\}
 = \bbrakket{\mathds{1}}{\rho^{(n)}(t)}
 \label{eq:prob}
 .
\end{equation}
To solve  Eqs.~(\ref{eq:n_alg}) we introduce a ``counting field'' $\chi$ via discrete Fourier transform $\kket{\rho(\chi ; t)} \equiv \sum_{n}{\kket{\rho^{(n)}}e^{in\chi}}$, to yield the following $\chi$-resolved master equation
\beq
  \cfrac{\mathrm{d}}{\mathrm{d} t} \kket{\rho (\chi; t)} = \left[\mathcal{L}+\mathcal{J}(\chi)\right]\kket{\rho(\chi ; t)}
  ,
\label{eq:rho_chi}
\eeq
with 
\beq
  \mathcal{J}(\chi) = 
  (e^{i\chi}-1)\mathcal{J}^+ 
  +(e^{-i\chi}-1)\mathcal{J}^- 
  \label{eq:JchiBI}
  .
\eeq
Solution of \eq{eq:rho_chi} is then used to find the CGF 
\beq
  F(\chi ; t) 
  = 
  \ln  \sum_{n} P(n;t) e^{in\chi}
  =
  \ln \bbrakket{\mathds{1}}{\rho(\chi; t)}
  ,
\eeq
from which the number cumulants are obtained as
\begin{equation}
  \ew{n^k(t)}_c =\left. \frac{\partial^k }{\partial( i\chi)^k}F(\chi ; t)\right|_{\chi=0}.
\end{equation}
The probability $P(n;t)$ can then be obtained from the inverse Fourier transform of the moment generating function $G(\chi;t) = e^{F(\chi ; t)}$.

As shown in Appendix \ref{sec:app}, in the long-time limit the CGF can be written as the expansion 
\begin{align}
F(\chi ; t) =  t\bigg(& \aav{\mathcal{J}(\chi)}+\aav{\mathcal{J}(\chi)\mathcal{R}\mathcal{J}(\chi)} \nonumber \\
                           & +\aav{\mathcal{J}(\chi)\mathcal{R}\mathcal{J}(\chi)\mathcal{R}\mathcal{J}(\chi)}+ \cdots \bigg)
\label{eq:cgf}
,
\end{align}
where $\aav{\cdot} \equiv \aav{\mathds{1}|\cdot|\rho_0} $ denotes the expectation value in the steady state,  and $\mathcal{R}$ is the pseudo-inverse of the Liouvillian
$
 \mathcal{R}=-\left[\mathcal{Q}\mathcal{L}\mathcal{Q}\right]^{-1}
$
with $\mathcal{Q} \equiv \mathds{1}-\kket{\rho_0}\bbra{\widetilde{\rho_0}}$ the projector out of the null-space (steady state subspace) of Liouvillian $\mathcal{L}$.

We will also consider unidirectional counting in which we only count transitions from state $1$ to $N$ and ignore the backwards transitions.  The above analysis holds identically with the exception that the $\chi$-resolved jump operator of \eq{eq:JchiBI} is replaced by 
\beq
  \mathcal{J}(\chi) = 
  (e^{i\chi}-1)\mathcal{J}^+ 
  \label{eq:JchiUNI}
  .
\eeq

\section{Large-system-size expansion}
\label{sec:expansion}

We now turn to the main focus of this paper which is the ensemble of random Liouvillians in the limit of large system size $N$.  In this section we calculate several properties of the  system that do not depend on exact choice of jump-operators, namely, the stationary state $\kket{\rho_0}$,
the projectors onto and out of the stationary state $\mathcal{P}\equiv \kket{\rho_0}\bbra{\widetilde{\rho_0}}$ and $\mathcal{Q}$, and the pseudo-inverse $\mathcal{R}$.  These are the building blocks required in building the CGF according to Eq. (\ref{eq:cgf}).

The main idea in this section is to decompose the random Liouvillian in the ensemble-averaged matrix $\overline{\mathcal{L}}$ and the deviation
$\Delta  \equiv   \mathcal{L}-\overline{\mathcal{L}}$.  We can write down expressions for the quantities of interest in terms 
of power series in deviation matrix $\Delta$.  From knowledge of the properties of the ensemble-average Liouvillian
we are able to translate the power series in $\Delta$ into one in terms of $1/N$, allowing us to keep only the first few terms in the limit of large systems.

\subsection{Ensemble-averaged quantities}

The ensemble-average of an $N$-dimensional Liouvillian is given by
\begin{equation}
\overline{\mathcal{L}}=\overline{\Gamma}\left(\left(\begin{matrix}
                                              1      & \cdots & 1 \\
                                              \vdots & \ddots & \vdots\\
                                              1      & \cdots & 1\\
                                              \end{matrix}\right)-N\mathds{1}\right) \equiv \overline{\Gamma}\left(\mathfrak{T}-N\mathds{1}\right),
\end{equation}
which defines $\mathfrak{T}$ as the matrix that has unit entries at every position.
This average Liouvillian has a single zero eigenvalue,  $\overline{\lambda_0}=0$, which corresponds to the stationary state.
The associated left eigenvector is as in \eq{eq:TrDVec}: 
$\overline{\bbra{\widetilde{\rho_0}}} = \bbra{\widetilde{\rho_0}}$.
The ensemble-averaged stationary state is given by
\begin{equation}
  \overline{\kket{\rho_0}}=\frac{1}{N}\kket{\mathds{1}}
  \label{eq:rho_av_right}
  .
\end{equation}

The remaining $(N-1)$ eigenvalues are all degenerate with the value $\overline{\lambda_k}=-N\overline{\Gamma};~ k = 1,\ldots,N-1$. 
We choose the right eigenvectors:
\begin{equation}
  \overline{\kket{\rho_k}} = -\kket{e_1} + \kket{e_{k+1}},
\end{equation}
where $\kket{e_k}$ denotes the $k$-th unit vector in the chosen basis, such that 
$
  \rb{
    \overline{\mathcal{L}} + N\overline{\Gamma}
  } \overline{\kket{\rho_k}} = 0
$.
The corresponding left eigenvectors are determined by orthonormality as
\begin{equation}
\overline{\bbra{\widetilde{\rho_k}}} =\frac{N-1}{N}\bbra{e_{k+1}}-\frac{1}{N}\sum_{m=1 \atop m\neq k}^{N}{\bbra{e_m}}
\end{equation}
This eigensystem provides us the ensemble-averaged projectors:
\begin{align}
\overline{\mathcal{P}}&=\overline{\kket{\rho_0}}\;\overline{\bbra{\widetilde{\rho_0}}}=\frac{1}{N}\mathfrak{T}, \\
\overline{\mathcal{Q}}&=\mathds{1}-\overline{\mathcal{P}}=-\frac{1}{N}\left(\mathfrak{T}-N\mathds{1}\right)
\end{align}
and pseudo-inverse:
\begin{align}
\overline{\mathcal{R}} & =-\sum\limits_{k=1}^{N-1}{\cfrac{\overline{\kket{\rho_k}}\;\overline{\bbra{\widetilde{\rho_k}}}}{\overline{\lambda_k}}} \nonumber \\
 & =\frac{N-1}{N^2\overline{\Gamma}}\sum_{k=1}^{N}{\kket{e_k}\bbra{e_k}}
                               -\frac{1}{N^2\overline{\Gamma}}{\sum_{k=1}^{N}{\sum_{m=1 \atop m\neq k}^{N}{\kket{e_k}\bbra{e_m}}}} \nonumber \\
 & =-\cfrac{1}{N^2\overline{\Gamma}}\left(\mathfrak{T}-N\mathds{1}\right).
\label{r_av}
\end{align}

\subsection{Expansion of the stationary state}
Let us write a member of our Liouvillian ensemble as 
$
  \mathcal{L}=
  \overline{\mathcal{L}}+\Delta
$, 
such that $\Delta$ describes the difference between the actual Liouvillian and the ensemble average.
The stationary state of $\mathcal{L}$ is then given by
\begin{equation}
 \left(\overline{\mathcal{L}}+\Delta\right)\kket{\rho_0}=0.
 \label{eq:me_pert}
\end{equation}
and we will look for a solution in terms of a power series, 
$\kket{\rho_0} = \sum_{k}{\delta\kket{\rho_0^{(k)}}}$,
where $\delta\kket{\rho_0^{(k)}}$ is of order $\Delta^k$.  Equating powers of $\Delta$ in Eq.~(\ref{eq:me_pert}) results in following recursion relation:
\begin{equation}
\overline{\mathcal{L}}\kket{\delta\rho_0^{(k)}} = \begin{cases}
                                                   0 & k=0,\\
                                                   -\Delta\kket{\delta\rho_0^{(k-1)}} & k \ge 1.
                                                  \end{cases}
\end{equation}
The zeroth-order term is therefore the ensemble average, $ \overline{\kket{\rho_0}}$ and the higher-order contributions, by iteration, are seen to be
\begin{equation}
\kket{\delta\rho_0^{(k)}}=-\left(\overline{\mathcal{Q}}\;\overline{\mathcal{L}}\;\overline{\mathcal{Q}}\right)^{-1}\Delta\kket{\delta\rho_0^{(k-1)}}
                         =\left(\overline{\mathcal{R}}\Delta\right)^k\overline{\kket{\rho_0}}.
\label{eq:rho(k)}
\end{equation}
Here we have employed to projectors $\overline{\mathcal{Q}}$ to remove the singular contributions of the Liouvillian.
Since, from \eq{eq:rho_av_right} and \eq{r_av}, we know that the elements of $\overline{\mathcal{R}}$ scale at worst as $N^{-1}$ and similarly for $\overline{\kket{\rho_0}}$, it follows that the stationary-state contributions have the $N$-dependence: 
\begin{equation}
  \kket{\delta\rho_0^{(k)}}_i \propto \cfrac{1}{N^{k+1}}
  + \text{higher-order terms}
  .
\end{equation}
Thus, the above expansion forms the basis for an expansion of the stationary quantities in terms of $1/N$.

\begin{figure}
 \includegraphics[width=\linewidth]{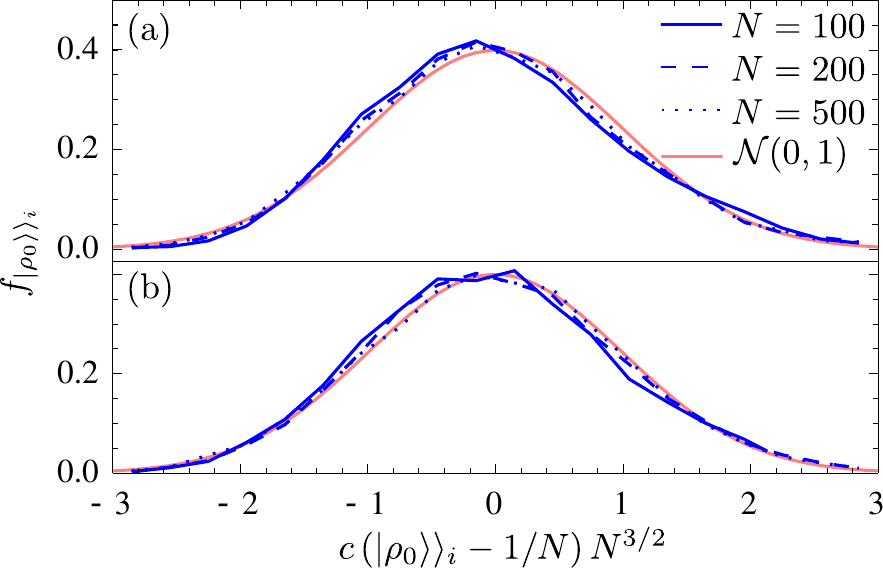}
 \caption{
  Probability density histograms for entries of the steady state vector for different asymmetric ensembles with $10^4$ matrices.
  Tunnel rates were chosen randomly from: {\bf (a)} the exponential distribution with $\overline{\Gamma}=1$; and {\bf (b)} the gamma-distribution with $\alpha=2$, $\beta=1$.
  The results are shifted by the mean and scaled with factors $N^{3/2}$ and $c=\overline{\Gamma}\left[2(\overline{\Gamma^2}-\overline{\Gamma}^2)\right]^{-1/2}$ as in the analytic distribution of Eq. (\ref{eq:rho_pdf}) (also shown).
 }
 \label{fig:rho_num}
\end{figure}

The lowest-order correction to the ensemble average stationary state is then:
\begin{equation}
\kket{\delta\rho_0^{(1)}}_i=\cfrac{1}{\overline{\Gamma}N^2}\sum_{j\neq i}{\left(\Gamma_{ij}-\Gamma_{ji}\right)}.
\label{eq:corr_rho}
\end{equation}
For symmetric matrices this contribution, and by \eq{eq:rho(k)} all higher ones, vanish, such that the stationary state is exactly equal to ensemble average
\begin{equation}
  \kket{\rho_0}=\overline{\kket{\rho_0}}=\frac{1}{N}\kket{\mathds{1}},
\end{equation}
which is consistent with the fact
that the relation $\kket{\rho_0} \propto \left( \bbra{\widetilde{\rho_0}}^*\right)^T$ is always valid for symmetric matrices.
In case of asymmetric ensembles, \eq{eq:corr_rho} allows us to find an approximation for the probability density function for the elements of the stationary state vector $\kket{\rho_0} \approx \overline{\kket{\rho_0}}+\kket{\delta\rho_0^{(1)}}$.
Since the tunnel rates are stochastically independent of one other, the summation in  Eq.~(\ref{eq:corr_rho}) can be evaluated according to the  central limit theorem.
This results in the distribution  for the elements of the stationary state vector:
\begin{equation}
\label{eq:rho_pdf}
 f_{\kket{\rho_0}_i}\left(x\right)\approx\mathcal{N}\left(\cfrac{1}{N},\cfrac{2\left(\overline{\Gamma^2}-\overline{\Gamma}^2\right)}{N^3\overline{\Gamma}^2}\right),
\end{equation}
where $\mathcal{N}(\mu ,\sigma^2)$ denote the normal distribution with mean $\mu$ and variance $\sigma^2$.
Numerical results for the distribution of steady-state vector entries are shown in Fig. \ref{fig:rho_num} for $N=100,200, 500$ and both exponentially and gamma-distributed transition rates. This results show good agreement with the analytically determined distribution function.

\subsection{Expansion of projectors}

Let us write $\mathcal{P} = \overline{\mathcal{P}} + \delta \mathcal{P}$ and similarly for $\mathcal{Q}$.  
From the resolution of the identity
$
 \mathcal{Q}+\mathcal{P}=\overline{\mathcal{Q}}+\overline{\mathcal{P}}=\mathds{1}
$
it follows that
$\delta\mathcal{Q}+\delta\mathcal{P}=0$.  The quantities $\delta\mathcal{P}$ and $\delta\mathcal{Q}$ can therefore be expressed in terms of the above expansion for the stationary state as
\beq
  \delta\mathcal{Q}
  =-\delta\mathcal{P}
  &=&
  \kket{\overline{\rho}_0}\bbra{\widetilde{\rho_0}}-\kket{\rho_0}\bbra{\widetilde{\rho_0}} 
  \nonumber\\
  &=&- \sum_{k \neq 0} \kket{\delta \rho_0^{(k)}}\bbra{\mathds{1}}
  \label{corr_q}
  .
\eeq
The scaling properties of the terms in this expansion are therefore the same as for the steady state vector itself.


\subsection{Expansion of the pseudo-inverse}
With 
$
  \delta\left(\mathcal{Q}\mathcal{L}\mathcal{Q}\right)
  \equiv
  \mathcal{Q}\mathcal{L}\mathcal{Q}
  -
  \overline{\mathcal{Q}}\,\overline{\mathcal{L}}\,\overline{\mathcal{Q}}
$
the pseudo-inverse can be expanded as
\begin{align}
 \mathcal{R} &=-\left[\overline{\mathcal{Q}}\,\overline{\mathcal{L}}\,\overline{\mathcal{Q}}+\delta\left(\mathcal{Q}\mathcal{L}\mathcal{Q}\right)\right]^{-1} \nonumber \\
             &=\overline{\mathcal{R}}\sum_{k=0}^{\infty}{\left(\delta\left(\mathcal{Q}\mathcal{L}\mathcal{Q}\right)\overline{\mathcal{R}}\right)^k}.
\label{eq:r_sum}
\end{align}
We first find an expression for $\delta\left(\mathcal{Q}\mathcal{L}\mathcal{Q}\right)$ using
\begin{align}
\label{eq:qlq}
\mathcal{Q}\mathcal{L}\mathcal{Q} & =\left(\overline{\mathcal{Q}}+\delta\mathcal{Q}\right)\mathcal{L}\left(\overline{\mathcal{Q}}+\delta\mathcal{Q}\right) \nonumber \\
                                  & =\overline{\mathcal{Q}}\,\overline{\mathcal{L}}\,\overline{\mathcal{Q}}
                                    +\overline{\mathcal{Q}}\Delta\overline{\mathcal{Q}}
                                    +\overline{\mathcal{Q}}\,\mathcal{L}\,\delta\mathcal{Q}
                                    +\delta\mathcal{Q}\,\mathcal{L} \left(\overline{\mathcal{Q}}+\delta\mathcal{Q}\right)
                                    .
\end{align}
\begin{figure}[t]
 \includegraphics[width=\linewidth]{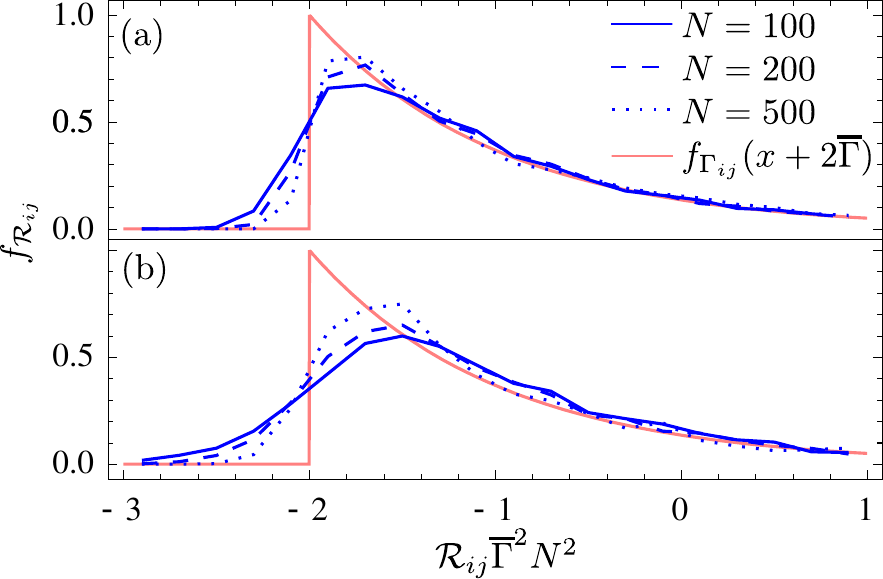}
 \caption{
  Probability density histograms for non-diagonal entries of the pseudo-inverse for $10^4$ matrices from {\bf (a)} 
  symmetric and {\bf (b)} asymmetric ensembles.
  Tunnel rates were chosen randomly from the exponential distribution with $\overline{\Gamma}=1$.
  The analytical approximation of Eq.~(\ref{eq:r_pdf}) is also shown.
 }
 \label{fig:rho_num_exp}
\end{figure}
With Eqs.~(\ref{r_av}), (\ref{corr_q}) and (\ref{eq:L})
the relation $\delta\mathcal{Q}\,\mathcal{L}=\delta\mathcal{Q}\,\overline{\mathcal{R}}=0$ can be shown easily.
Consequently, the series for the pseudo-inverse simplifies to
\begin{equation}
 \mathcal{R} =\sum_{k=0}^{\infty}{\left(\overline{\mathcal{R}}\Delta\right)^k\overline{\mathcal{R}}}.
\end{equation}
Leading terms in higher orders corrections of $\mathcal{R}$ are scaled same as the steady state vector entries:
\begin{equation}
 \delta\mathcal{R}_{ij}^{(k)} \propto \cfrac{1}{N^{k+1}}, \quad k > 0
 .
\end{equation}
So we get the first correction of the pseudo-inverse elements:
\beq
    \delta\mathcal{R}_{ii}^{(1)}= 
      \sum_{m \neq i}
      -\frac{\left(\Gamma_{im}-\overline{\Gamma}\right)}
      {N^2\overline{\Gamma}^2} 
      +
      \frac{\left(\Gamma_{im}+\Gamma_{mi}-2 \overline{\Gamma}\right)}{N^3\overline{\Gamma}^2}
      ,
 \label{eq:r_diag}
\eeq
on the diagonal and 
\beq
    \delta\mathcal{R}_{ij}^{(1)}
    =
    \frac{\left(\Gamma_{ij}-\overline{\Gamma}\right)}{N^2\overline{\Gamma}^2} 
    +
    \sum_{m\neq i}
    \frac{\left(\Gamma_{mi}-\Gamma_{im}\right)}{N^3\overline{\Gamma}^2}
    ,
 \label{eq:r_offdiag}
\eeq
for the off-diagonal elements.

To lowest-order in $1/N$, an off-diagonal element of $\mathcal{R}$ therefore reads
\begin{equation}
  \mathcal{R}_{ij} 
  \approx 
  \frac{\Gamma_{ij}-2\overline{\Gamma}}{N^2\overline{\Gamma}^2}
  \label{eq:r_corr}
  ,
\end{equation}
which implies that the off-diagonal elements of the pseudo-inverse are distributed according to the same distribution as the rates, albeit  rescaled and shifted:
\begin{equation}
  f_{\mathcal{R}_{ij}}(x) \approx N^2\overline{\Gamma}^2
  f_{\Gamma_{ij}}\left(N^2\overline{\Gamma}^2 x - 2\overline{\Gamma}\right)
  \label{eq:r_pdf}
  .
\end{equation}
%

\begin{figure}
 \includegraphics[width=\linewidth]{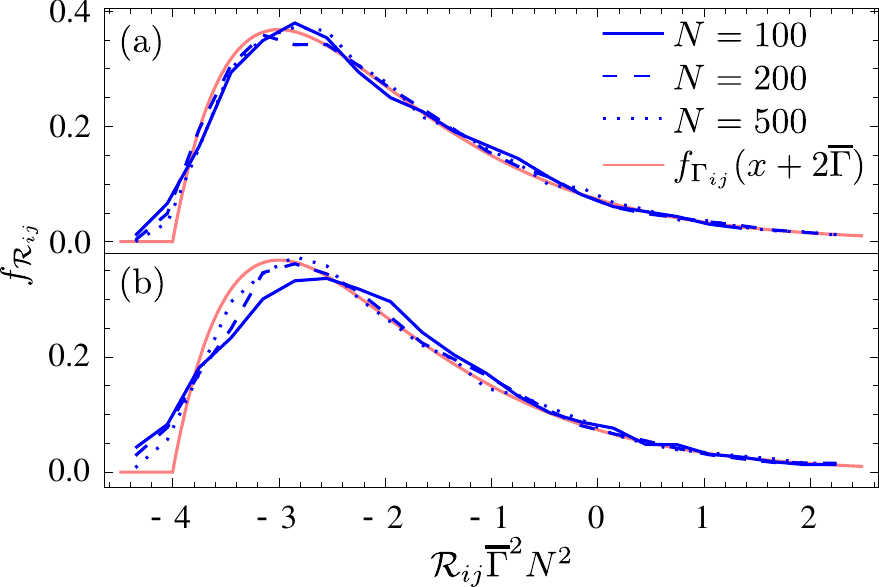}
 \caption{
  As Fig.~\ref{fig:rho_num_exp} but with rates distributed according to the gamma-distribution with $\alpha=2$, $\beta=1$.
  \label{fig:rho_num_gamma}
 }
\end{figure}

Figs.~\ref{fig:rho_num_exp} and \ref{fig:rho_num_gamma} show distributions of 
the pseudoinverse elements determined numerically.  Good agreement between numerical results and analytical approximation is observed.
Convergence to the ideal curve is slower in case of the exponential distribution due to its abrupt discontinuity.

\section{Ensemble-averaged CGFs}
\label{sec:CGFs}
With the quantities calculated in the previous section, we are in a position to calculate ensemble-averaged CGFs for both unidirectional and bidirectional transport up to an arbitrarily order in $1/N$. 
The scaling properties of all relevant quantities (at least of leading terms in $1/N$-expansion) are: $\mathcal{J}(\chi) \propto N^0$, $\kket{\rho_0} \propto N^{-1}$ and $\mathcal{R} \propto N^{-2}$.
We shall consider expansion of CGFs up to $N^{-3}$, so that we can approximate Eq.~(\ref{eq:cgf}) by
\begin{align}
  F(\chi ; t) & \approx  t\left( \aav{\mathcal{J}(\chi)}+\aav{\mathcal{J}(\chi)\mathcal{R}\mathcal{J}(\chi)} \right)
  \label{eq:cgf_approx}
  .
\end{align}

\subsection{Unidirectional transport}
\label{subsec:uni}
\subsubsection{Single-Channel Counting}
First we consider the case of counting events in only one direction
for which the $\chi$-dependent jump operator introduced in Sec. \ref{sec:model} is given by
 \begin{equation}
  \mathcal{J}\left(\chi\right)=\mathcal{J}^+\left(e^{i\chi}-1\right),
 \end{equation}
with $\mathcal{J}^+$ defined in Eq. (\ref{eq:j+}). The ensemble-averaged CGF is then
\beq
  \overline{F(\chi ; t)} 
  &=& 
  \overline{\aav{\mathcal{J}^+}} t 
  \left(e^{i\chi}-1\right)
  \nonumber\\
  &&
  +\overline{\aav{\mathcal{J}^+\mathcal{R}\mathcal{J}^+}} t
  \left(e^{i\chi}-1\right)^2
  \label{eq:cgf_uni} 
  .
\eeq
We first analyze $\overline{\aav{\mathcal{J}^+}}$ using the series expansion of the steady state:
\begin{equation}
 \overline{\aav{\mathcal{J}^+}} \approx \bbra{\mathds{1}}\overline{\mathcal{J}^+}\,\overline{\kket{\rho_0}}+\bbra{\mathds{1}}\overline{\mathcal{J}^+\kket{\delta \rho^{(1)}_0}} +\bbra{\mathds{1}}\overline{\mathcal{J}^+\kket{\delta \rho^{(2)}_0}}.
\label{eq:j+_exp}
 \end{equation}
Since $\kket{\delta \rho^{(k )}_0}$ vanish for all $k \neq 0$ in case of symmetric ensembles, we get $\overline{\aav{\mathcal{J}^+}}_{\mathrm{sym}}=\overline{\Gamma}/N$.
For asymmetric ensembles evaluation of Eq.~(\ref{eq:j+_exp}) leads to
\begin{equation}
 \overline{\aav{\mathcal{J}^+}}_{\mathrm{asym}} \approx \cfrac{\overline{\Gamma}}{N} +\cfrac{\overline{\Gamma^2}-\overline{\Gamma}^2}{N^2\overline{\Gamma}}+\cfrac{\overline{\Gamma^3}+\overline{\Gamma}^3}{N^3\overline{\Gamma}^2}
\end{equation}
In order to calculate $\aav{\mathcal{J^+RJ^+}}$ we proceed as above:
\begin{align}
 &\overline{\aav{\mathcal{J^+RJ^+}}}= \nonumber \\ &\bbra{\mathds{1}}\mathcal{\overline{J^+\overline{R}J^+}}\overline{\kket{\rho_0}} 
                          + \bbra{\mathds{1}}\overline{\mathcal{J^+}\delta \mathcal{R}^{(1)}\mathcal{J^+}}\:\overline{\kket{\rho_0}} \nonumber \\
                         +&\bbra{\mathds{1}}\overline{\mathcal{J^+}\overline{\mathcal{R}}\mathcal{J^+}\kket{\delta \rho_0^{(1)}}}
                          +\bbra{\mathds{1}}\overline{\mathcal{J^+}\delta\mathcal{R}^{(1)}\mathcal{J^+}\kket{\delta \rho_0^{(1)}}}.
\end{align}
The last two terms in this equation scale with $N^{-4}$ and so will be neglected. The remain terms give:
\begin{align}
 \overline{\aav{\mathcal{J^+RJ^+}}}_{\mathrm{sym}} & = \frac{\overline{\Gamma^3}-2\overline{\Gamma^2}\,\overline{\Gamma}}{N^3\overline{\Gamma}^2},  \\
 \overline{\aav{\mathcal{J^+RJ^+}}}_{\mathrm{asym}} & = \frac{-\overline{\Gamma^2}}{N^3\overline{\Gamma}}.
 \label{eq:jrj_asym}
\end{align}
Consequently, we get different CGFs according to symmetry properties of the matrix ensemble:
 \begin{align}
   \overline{F(\chi ; t)}_{\mathrm{sym}} & =  t \left[\frac{\overline{\Gamma}}{N}\left(e^{i\chi}-1\right)+\frac{\overline{\Gamma^3}-2\overline{\Gamma^2}\,\overline{\Gamma}}{N^3\overline{\Gamma}^2}\left(e^{i\chi}-1\right)^2\right],\\
   \overline{F(\chi ; t)}_{\mathrm{asym}} & =  t\left[ \left(\cfrac{\overline{\Gamma}}{N} +\cfrac{\overline{\Gamma^2}-\overline{\Gamma}^2}{N^2\overline{\Gamma}}+\cfrac{\overline{\Gamma^3}+\overline{\Gamma}^3}{N^3\overline{\Gamma}^2} \right)\left(e^{i\chi}-1\right)\right. \nonumber \\
                                          &             \left.-\frac{\overline{\Gamma^2}}{N^3\overline{\Gamma}}\left(e^{i\chi}-1\right)^2 \right].
 \end{align}
%
\begin{figure}
 \includegraphics[width=\linewidth]{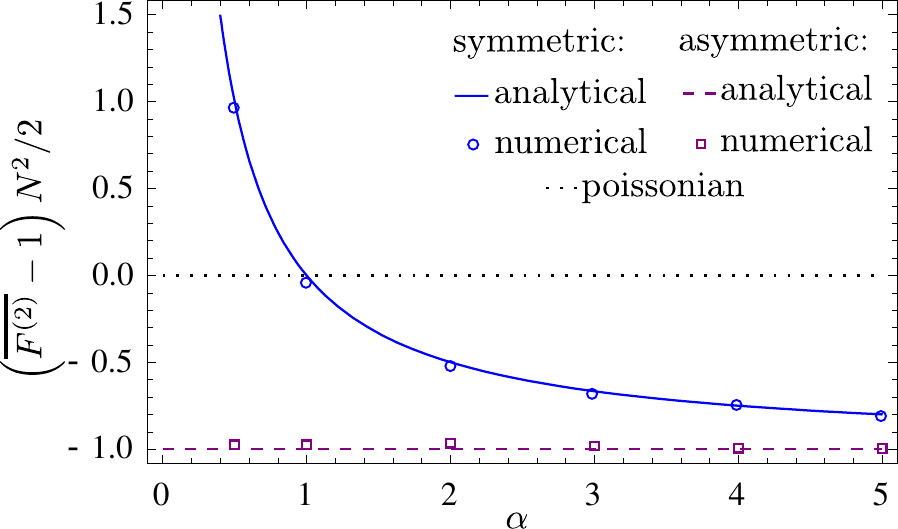}
 \caption{
	  Ensemble-averaged Fano factors for both symmetric and asymmetric ensembles with tunnel rates chosen from the gamma-distribution with parameters $\alpha$ and $\beta$.  
	  We plot the scaled deviations from the Poisson value of unity.
	  The continuous lines shows analytic results of Eq.~(\ref{eq:fano_asym}) and Eq.~(\ref{eq:fano_cases}), whereas the symbols show numerical averages from $10^4$ matrices of size $N=50$          
          }
  \label{fig:fano}
\end{figure}
%
To lowest order in $1/N$, the counting statistics of these random systems is Poissonian with a 
characteristic rate $\overline{\Gamma}/N$, i.e. the mean rate of the individual rate distribution scaled by the inverse system size.
This makes sense in the network picture.  When the system is very large, having just left the counting vertex,
it takes the system such a long time to wander around and finally return to it, that counting events are essentially independent 
of one another, and this is the basis of the Poisson distribution.
Considering the first correction to this Poissonian limit, which occurs at order $N^{-2}$, we see a difference between symmetric and asymmetric Liouvillians.  In both cases, the FCS remain Poissonian, but  in the asymmetric case only, the rate obtains a correction that depends on the second cumulant of the rate distribution. At this order, the FCS for the symmetric case remains unaltered.
Finally, at order $N^{-3}$ we see deviations from Poissonian statistics, which are a signature of correlations between counting events.
To get a handle on these correlations, let us define the (second-order) Fano factor as the ratio of the second to first cumulants:
$
  F^{(2)} \equiv \ew{n^2(t)}_c /  \ew{n(t)}_c
$. Its ensemble-average is than given by $\overline{F^{(2)}}=1+2\overline{\aav{\mathcal{JRJ}}/\aav{\mathcal{J}}}$.
Calculating just the leading order terms we obtain in the symmetric case
\beq
 F^{(2)}_{\mathrm{sym}} &=1-\cfrac{2}{N^2}\left( 2-\cfrac{\overline{\Gamma^2}}{\overline{\Gamma}^2}\right)
 ,
\eeq
and in the asymmetric case
\beq
 F^{(2)}_{\mathrm{asym}} &=1-\cfrac{2}{N^2}.
 \label{eq:fano_asym}
\eeq
Here the deviation from Poissonian behaviour $F^{(2)}=1$ is clearly seen as correction scaling as $N^{-2}$,
which differs depending on the symmetry of the problem.
Interestingly, in the asymmetric case, the leading order correction is independent of the $\Gamma_{ij}$-distribution, whereas in symmetric case, it
depends on the first two moments.
Moreover, in the asymmetric case, the correction must always be negative and the statistics
correspondingly subPoissonian (see Ref.~\cite{PhysRevB.85.165417} for a discussion on the significance of this point). In contrast, in the symmetric case,
the correction can be of either sign and the statistics
correspondingly sub- or super-Poissonian.

We calculate Fano factors for both exponential and gamma-distribution for the symmetric case
\begin{equation}
\overline{F^{(2)}}=  
 \begin{cases}
  1, & \text{exp. distr.,}\\
  1-\cfrac{2}{N^2}\left(1- \cfrac{1}{\alpha}\right), & \text{gamma-distr.}
 \end{cases}
\label{eq:fano_cases}
\end{equation}
On the one hand, we see that in case of exponential distribution the behaviour is always Poissonian.
On the other hand, we also clearly see the sign change in case of gamma-distribution at $\alpha=1$, 
which corresponds to exponential distribution with $\overline{\Gamma}=\beta$, and thus a change from superPoissonian
to subPoissonian behaviour with increasing alpha.

Fig.~\ref{fig:fano} shows numerical results for the Fano factor of both symmetric and asymmetric ensembles with gamma-distributed tunnel rates.
We see a good match with analytical results given in Eq.~(\ref{eq:fano_asym}) and Eq.~(\ref{eq:fano_cases}).

\subsubsection{Multi-Channel Counting}
In case of unidirectional counting of events we can also consider more than just transitions between two states. Here
we assume that we count transitions from a arbitrary set of nodes $\{k\}$ to state $1$. We also assume the cardinality $K$ 
of this set to be much smaller than $N$. The $\chi$-dependent jump operator in this case is given by a superposition 
of jump operators for the individual transitions:
\begin{equation}
 \mathcal{J}^+(\chi) = \sum_{l \in \{k\}}\mathcal{J}_l^+ (e^{i\chi}-1)
 ,
\end{equation}
where the sum is over the relevant input nodes, and where the individual jump operators $\mathcal{J}_l^+ $ have matrix elements
\begin{equation}
  \rb{\mathcal{J}_l^+}_{ij} = \Gamma_{ij}\delta_{1i}\delta_{lj}
  .
\end{equation}
With these forms, the important quantities in Eq. (\ref{eq:cgf_uni}) read
\begin{align}
 \overline{\aav{\mathcal{J}^+}} &=\sum_{l\in \{k\}}{\overline{\aav{\mathcal{J}_l^+}}}=K\overline{\aav{\mathcal{J}_l^+}};
 \\
 \overline{\aav{\mathcal{J}^+\mathcal{R}\mathcal{J}^+}} 
 &= K
 \left[
   \overline{\aav{\mathcal{J}_l^+\mathcal{R}\mathcal{J}_l^+}}
 \right.
 \nonumber\\
 &~~~~~~~~~
 \left.
   +\left(K-1\right)\overline{\aav{\mathcal{J}_l^+\mathcal{R}\mathcal{J}_{l'}^+}}
 \right]
 \label{eq:jrj_k_jump}
\end{align}
for arbitrary $l, l' \in  \{k\}$ with $l\neq l'$.
Taking the ensemble average, the first summand in Eq. (\ref{eq:jrj_k_jump}) have already been found in the single jump case and second evaluates as
\begin{align}
 \overline{\aav{\mathcal{J}_l^+\mathcal{R}\mathcal{J}_{l'}^+}}_{\mathrm{sym}} &= \cfrac{\overline{\Gamma^2}-2\overline{\Gamma}^2}{N^3\overline{\Gamma}},\\
 \overline{\aav{\mathcal{J}_l^+\mathcal{R}\mathcal{J}_{l'}^+}}_{\mathrm{asym}} &= -\cfrac{\overline{\Gamma}}{N^3}
 .
\end{align}
The resulting  $K$-dependent CGFs for symmetric and asymmetric rates read:
\begin{multline}
   \overline{F(\chi ; t; K)}_{\mathrm{sym}}  =  t \Bigg[K\frac{\overline{\Gamma}}{N}\left(e^{i\chi}-1\right)+\\ \frac{K}{N^3}\left(-2(K-1)\overline{\Gamma}+(K-3)\frac{\overline{\Gamma^2}}{\overline{\Gamma}}+\frac{\overline{\Gamma^3}}{\overline{\Gamma}^2}\right)\left(e^{i\chi}-1\right)^2\Bigg],
\label{eq:mult_sym}
   \end{multline}
and
   \begin{multline}
   \overline{F(\chi ; t; K)}_{\mathrm{asym}}  =\\  t\left[ K\left(\cfrac{\overline{\Gamma}}{N} +\cfrac{\overline{\Gamma^2}-\overline{\Gamma}^2}{N^2\overline{\Gamma}}+\cfrac{\overline{\Gamma^3}+\overline{\Gamma}^3}{N^3\overline{\Gamma}^2} \right)\left(e^{i\chi}-1\right)\right.  \\
                                                \left.-\frac{K}{N^3}\left( (K-1)\overline{\Gamma}+\frac{\overline{\Gamma^2}}{\overline{\Gamma}}\right)\left(e^{i\chi}-1\right)^2 \right].
\label{eq:mult_asym}
                                                \end{multline}

\subsection{Bidirectional transport}
\label{subsec:bi}
In the case of bidirectional counting at a single vertex, the $\chi$-dependent jump operator is given by
\begin{equation}
 \mathcal{J}\left(\chi\right)=\mathcal{J}^+\left(e^{i\chi}-1\right)+\mathcal{J}^-\left(e^{-i\chi}-1\right).
\end{equation}
Taking the ensemble average of the first term in Eq. (\ref{eq:cgf_approx}) we get
\begin{equation}
 \overline{\aav{\mathcal{J}\left(\chi\right)}}=2\overline{\aav{\mathcal{J}^+}}\left(\mathrm{cosh}\left(i\chi\right)-1\right).
\end{equation}
Expanding the second term in Eq.~(\ref{eq:cgf_approx}) in terms of $\mathcal{J}^\pm$ results in terms:
$\overline{\aav{\mathcal{J^+RJ^+}}}$ and $\overline{\aav{\mathcal{J^-RJ^-}}}$, which are equal to each other and to the 
corresponding term in the unidirectional case; as well as the cross terms
$\overline{\aav{\mathcal{J^+RJ^-}}}$ and $\overline{\aav{\mathcal{J^-RJ^+}}}$, which become equal in the average and evaluate as
\begin{align}
 \overline{\aav{\mathcal{J^+RJ^-}}}_{\mathrm{sym}}&=\frac{\overline{\Gamma^2}}{N^2\overline{\Gamma}}-\frac{\overline{\Gamma^3}}{N^3\overline{\Gamma}^2};
 \\
 \overline{\aav{\mathcal{J^+RJ^-}}}_{\mathrm{asym}}&=\frac{\overline{\Gamma}}{N^2}-\frac{\overline{\Gamma^2}}{N^3\overline{\Gamma}},
\end{align}
for the two different symmetry cases.
With these results, we find ensemble-averaged CGFs  in case of bidirectional counting:
\begin{multline}
 \overline{F(\chi ; t)}_{\mathrm{sym}}  = \\ t \left[\left(\frac{2\overline{\Gamma}}{N}-\frac{4\overline{\Gamma^2}}{N^2 \overline{\Gamma}}+\frac{4\overline{\Gamma^3}}{N^3 \overline{\Gamma}^2}\right)\left(\mathrm{cosh}\left(i\chi\right)-1\right)\right.\\
  +\left(\frac{\overline{\Gamma^3}}{N^3\overline{\Gamma}^2}-\frac{2\overline{\Gamma^2}}{N^3\overline{\Gamma}}\right)\left( \left(e^{i\chi}-1\right)^2+\left(e^{-i\chi}-1\right)^2\right)\Bigg],
\end{multline}
and
\begin{multline}
 \overline{F(\chi ; t)}_{\mathrm{asym}}  =\\  t \left[2\left(\frac{\overline{\Gamma}}{N}-\frac{\overline{\Gamma^2}-3\overline{\Gamma}^2}{N^2\overline{\Gamma}}+\frac{\overline{\Gamma}^3+\overline{\Gamma}\overline{\Gamma^2}+\overline{\Gamma^3}}{N^3\overline{\Gamma}^2}\right) \left(\mathrm{cosh}\left(i\chi\right)-1\right)\right.\\
   -\frac{1}{N^3}\frac{\overline{\Gamma^2}}{\overline{\Gamma}}\left( \left(e^{i\chi}-1\right)^2+\left(e^{-i\chi}-1\right)^2\right)\Bigg]
   .
\end{multline}
To lowest order in $N^{-1}$, both CGFs are identical and equal to that of a system with bidirectional Poisson processes with the equal rates $\overline{\Gamma}/N$.
This means that all odd cumulants vanish, and this property is maintained when higher order corrections are considered. 
Similarly to the unidrectional case, the two CGFs differ from one another at order $N^{-2}$ 
and deviation from Poisson (in this case bidirectinal Poisson) behaviour becomes first evident at order $N^{-3}$.

\section{Discussion \label{SEC:disc}}

We have presented a method for the evaluation of ensemble-averaged FCS  large systems whose time evolution is described by a ME with a random Liouvillian.
The method is based upon a series expansion of the relevant system quantities based on the deviation of the system from its ensemble average.
Our approach therefore has the character of a mean-field solution and is enabled by the all-to-all structure of the network.
The series expansion for the CGF can be rephrased in terms of an expansion in terms of $1/N$.  At lowest order with simple jump operators, the FCS just looks Poisson with rescaled rate.  The first  differences between symmetric and asymmetric Liouvillians appear at order $N^{-2}$.  Finally,  first at order $N^{-3}$ do deviations from Poisson behaviour appear. This is more clearly manifest in the Fano factor for unidrectional transport, where only symmetric systems can give rise to  super-Poissonian behaviour.

\begin{acknowledgments}
We are grateful to T.~Brandes for useful discussions.  This work was supported by DFG Grant GRK 1558.
\end{acknowledgments}

\begin{appendix}
\section{CGF in the long time limit}
\label{sec:app}
In this appendix we derive the CGF of \eq{eq:n_alg}.  The approach is realated to the work discussed in Refs~\onlinecite{Flindt2008,Flindt2010}.
We start with the $\chi$-dependent master equation:
\begin{align}
\cfrac{\mathrm{d}}{\mathrm{d} t} \kket{\rho(\chi; t)}& = \left[\mathcal{L}+\mathcal{J}(\chi)\right]\kket{\rho(\chi ; t)}
.
\end{align}
Laplace transformation of this expression leads to
\begin{align}
\kket{\rho(\chi; z)}& =\left[z-\mathcal{L}(\chi)\right]^{-1}\kket{\rho} \nonumber \\
             & =\sum_{n=0}^{\infty}{\Omega_0(z)\left(\mathcal{J}(\chi)\Omega_0(z)\right)^n\kket{\rho}},
             \label{eq:rho_chi_z}
\end{align}
with the propagator $\Omega_0(z) \equiv \left[z-\mathcal{L}\right]^{-1}$ and the initial distribution $\kket{\rho(t=0)}$.
The moment generating function $G(\chi ; t)$ and CGF  $G(\chi ; t)$ are then given by
\begin{equation}
G(\chi ; t)=e^{F(\chi ; t)} \equiv \sum_{n}{P(n ; t)e^{in\chi}}= \mathrm{Tr}\left\{ \kket{\rho(\chi;t)}\right\}.
\label{eq:f}
\end{equation}

The right side of Eq. (\ref{eq:f}) can be analysed in the frequency domain using Eq. (\ref{eq:prob}) and (\ref{eq:rho_chi_z}).
For the long-time behaviour of the system
the choice of the initial condition becomes irrelevant and we can pick out the steady state distribution $\kket{\rho(t=0)} \to \kket{\rho_0}$.
We therefore obtain
\begin{align}
G(\chi ; z) & =\aav{\sum_{n=0}^{\infty}{\Omega_0(z)\left(\mathcal{J}(\chi)\Omega_0(z)\right)^n}},
\label{eq:tr_rho}
\end{align}
with the expectation as defined in Sec.~\ref{sec:model}.

We then rewrite the propagator $\Omega_0$ with help of spectral
decomposition of $\mathcal{L}$. We decompose it into singular and non-singular parts
\begin{equation}
  \left[z-\mathcal{L}\right]^{-1}=\sum\limits_{k=0}^{N-1}{\frac{\kket{\rho_k}\bbra{\rho_k}}{z-\lambda_k}}
  =
  \frac{\mathcal{P}}{z}
  +
  \mathcal{R}(z)
  .
\end{equation}

Since we are interested in the long-time limit, we can use the final value theorem and study the zero frequency behaviour.  Taking the $z \rightarrow 0$ limit for the pseudo-inverse,  $\mathcal{R} = \mathcal{R}(z=0)$, we rewrite Eq. (\ref{eq:tr_rho}) as
\begin{align}
G(\chi ; z) &=\frac{1}{z}+\frac{1}{z^2}\aav{\mathcal{J}(\chi)} \nonumber \\
                                       & + \frac{1}{z^3}\aav{\mathcal{J}(\chi)}^2 +\frac{1}{z^2}\aav{\mathcal{J}(\chi)\mathcal{R}\mathcal{J}(\chi)} \nonumber \\
                                       & + \frac{1}{z^4}\aav{\mathcal{J}(\chi)}^3 +\frac{2}{z^3}\aav{\mathcal{J}(\chi)}\aav{\mathcal{J}(\chi)\mathcal{R}\mathcal{J}(\chi)}\nonumber \\
                                       & +\frac{1}{z^2}\aav{\mathcal{J}(\chi)\mathcal{R}\mathcal{J}(\chi)\mathcal{R}\mathcal{J}(\chi)} + \cdots  
\end{align}
Performing the inverse Laplace transform we get
\begin{align}
&\hspace{-.2cm} G(\chi ; t)=\\ \nonumber
 1 & + t\bigg( \aav{\mathcal{J}(\chi)}+\aav{\mathcal{J}(\chi)\mathcal{R}\mathcal{J}(\chi)} \nonumber \\
 &\quad \quad \quad  +\aav{\mathcal{J}(\chi)\mathcal{R}\mathcal{J}(\chi)\mathcal{R}\mathcal{J}(\chi)}+ \cdots \bigg) \nonumber\\
 &+t^2\left(\frac{1}{2}\aav{\mathcal{J}(\chi)}^2 + \aav{\mathcal{J}(\chi)}\aav{\mathcal{J}(\chi)\mathcal{R}\mathcal{J}(\chi)} + \cdots \right) \nonumber \\
 &+t^3\left(\frac{1}{6}\aav{\mathcal{J}(\chi)}^3+ \cdots \right) \nonumber \\
 & \hspace{.8cm} \equiv 1+ \widetilde{G}(\chi ; t),
\label{eq:e^f}.
 \end{align}

The CGF can be obtained from Eq.~(\ref{eq:e^f}) with $ F(\chi ; t)= \ln G(\chi ; t)$.
From conservation of probability we learn that
$\left. G(\chi ; t)\right|_{\chi=0}= \sum_{n}{P(n ; t)}\stackrel{\mathrm{!}}=1$, and hence,
$\left. \widetilde{G}(\chi ; t)\right|_{\chi=0}=0$.

At this point we can use the power series expansion for the logarithm around one
\begin{align}
 F(\chi ; t)& = \ln G(\chi ; t)=\ln \left( 1+ \widetilde{G}(\chi ; t) \right) \nonumber \\
            & =-\sum \frac{(-\widetilde{G}(\chi ; t))^{k+1}}{k+1},
\label{eq:ln_exp}
\end{align}\baselineskip0pt
since only the behaviour of the CGF around $\chi=0$  is interesting.

Combining Eq. (\ref{eq:e^f}) and Eq. (\ref{eq:ln_exp}) we get
\begin{align}
F(\chi ; t) =  t\bigg(& \aav{\mathcal{J}(\chi)}+\aav{\mathcal{J}(\chi)\mathcal{R}\mathcal{J}(\chi)} \nonumber \\
                           & +\aav{\mathcal{J}(\chi)\mathcal{R}\mathcal{J}(\chi)\mathcal{R}\mathcal{J}(\chi)}+ \cdots \bigg).
\end{align}

\end{appendix}
\end{document}